\documentclass[a4paper,11pt]{article}

\usepackage{CJK,CJKnumb,indentfirst,graphicx,bm,latexsym,amsmath,multicol,subfigure,titletoc,lscape,braket,amssymb,float,caption2}
\usepackage[bf,small,indentafter,pagestyles]{titlesec}
\usepackage[perpage,symbol]{footmisc}
\usepackage[numbers,sort&compress]{natbib}
\usepackage[hyperfootnotes=false]{hyperref}

\renewcommand{\citet}[1]{\textsuperscript{\cite{#1}}}

\def \bea {\begin{eqnarray}}
\def \eea {\end{eqnarray}}
\def \be {\begin{equation}}
\def \ee {\end{equation}}
\hypersetup{colorlinks, citecolor=red, urlcolor=blue}

\begin{document}
\begin{CJK*}{UTF8}{gbsn}

\title{\Large \textbf{Anisotropy in Inflation with Non-minimal Coupling}}
\author{\small Bin Chen$^{1,2,3}$\footnote{Email address: bchen01@pku.edu.cn} and Zhuang-wei Jin$^1$\footnote{Email address: jinzw@pku.edu.cn}}

\date{}

\maketitle


\begin{center}
\hspace{-1cm}\textit{\small $^1$Department of Physics and State Key Laboratory of Nuclear Physics and Technology,\\Peking University, 5 Yiheyuan Rd, Beijing 100871, P. R. China}\\
\mbox{\textit{\small $^2$Collaborative Innovation Center of Quantum Matter, 5 YiheyuanRd, Beijing 100871, P. R. China}}\\
\mbox{\textit{\small $^3$Center for High Energy Physics, Peking University, 5 Yiheyuan Rd, Beijing 100871, P. R. China}}
\end{center}

\vspace{1em}
\abstract{}
We study a new anisotropic inflation model, with an inflaton field nonminimally coupled with the gravity and a vector field. We find that the anisotropic attractor solution exists not only in the weak curvature coupling limit, but more interestingly in  the strong  curvature coupling limit as well. We show that in the strong curvature coupling limit, the contribution from the anisotropy is greatly suppressed.

\newpage
\section{Introduction}

Inflation provides a natural solution to several notorious puzzles in the standard Big Bang scenario, including the flatness problem, the horizon problem and the monopole problem. More importantly this mechanism sets the initial conditions for cosmological perturbations and predicts a nearly scale invariant power spectrum as well\citet{dodelson,baumann,guth,linde,brandenberger}. Thus far, it has been widely accepted. Despite its great success, there have been unsettled issues since its proposal. Among them, the origin of the inflaton which drives the inflation is the most outstanding one.  The most economical choice seems to  identify the only scalar field in Standard Model, the Higgs field, with the inflaton field\citet{higgs1,higgs2,higgs3,higgs4,higgs5}, especially after the discovery of LHC in 2012\citet{lhc1,lhc2}.

However, such identification does not work well in the minimal coupling case because of the conflict between the constraints on the properties of Higgs boson and the observed isotropy of the CMB. If we still assume that the inflation potential is of quartic form, the self-coupling constant $\lambda$ in the quartic potential has to be unnaturally smaller than $10^{-12}$ in order to give the correct magnitude of the scalar power spectrum\citet{komatsu1,higgs1,komatsu2}. To solve this problem, Fakir and Unruh introduced a non-minimal coupling term $\xi\kappa^2\phi^2 R$ with a coupling constant $\xi$\citet{higgs5}. Then the potential in Einstein frame is exponentially flat when $\xi\kappa^2\phi^2 \gg 1$\citet{higgs2}, ensuring slow-roll conditions for the inflation. In that scenario, a reasonable $\lambda \sim 10^{-2}$ can be reached if $\xi \sim 10^3$. One remarkable feature in the non-minimal Higgs inflation is that it has a quite small tensor-to-scalar ratio $r \approx 0.002$\citet{komatsu1}, in disfavor of recent result $r \simeq 0.2$ given by BICEP2\citet{bicep2}. Various attempts have been made to raise the value of $r$, for example, by considering a running kinetic inflation\citet{higgs_after_bicep2_1} or tuning the top quark mass\citet{higgs_after_bicep2_2}.

On the other hand, the recent precise cosmological observations indicate small deviations from isotropy\citet{planck,wmap}. Thus it is well-motivated to propose the inflation models generating statistical anisotropy naturally. Stable anisotropic inflation models have been  found to give rise to statistical anisotropy\citet{soda1}. This so-called $f(\phi)F^2$ mechanism is realized by introducing a vector field coupled with the inflaton $\phi$. By choosing the coupling function $f(\phi)$ appropriately, there exists an attractor in the slow-roll phase representing anisotropic inflation, with anisotropy being of the order of the slow-roll parameter $\epsilon$. Specifically, besides the small correction to the power spectra, the non-vanishing cross-correlations between the scalar perturbation and the tensor perturbation are expected to have imprints on the CMB spectrum\citet{soda5}.

In this paper, we investigate the anisotropic inflation in the non-minimal Higgs inflation models. We introduce the coupling term $f(\phi)F^2$ between the $U(1)$ vector field and the inflaton, besides the nonminimal  curvature coupling term $\xi\kappa^2\phi^2 R$. In order to have nonvanishing anisotropy, the form of the function $f(\phi)$ should include an exponential factor.
We show that the anisotropic attractor solutions exist in both the weak and strong curvature coupling limits, by using  analytical and numerical methods.  In particular, we find that the anisotropy is greatly suppressed in the strong curvature coupling limit for quite general form of the coupling function $f(\phi)$.

The rest of the paper is organized as follows. In Sec. 2, we study the attractor behavior of anisotropic inflation in our model. We first  show the existence of the anisotropic attractor solution under appropriate approximation. We argue that the coupling function $f$ should include an exponential function in order to have an unignorable anisotropy. We moreover show the suppression of the anisotropy in the strong curvature coupling limit.  Finally we use numerical analysis to support the picture.  In Sec. 3, we make conclusion of the work.

\section{Anisotropic inflation}

Let us consider the following action:
$$S=\int d^4x \sqrt{-g} [\frac{1}{2\kappa^2}(1+\kappa^2\xi\phi^2)R-\frac{1}{2} g^{\mu\nu} \partial_\mu \phi \partial_\nu \phi-\frac{\lambda}{4}\phi^4-\frac{f^2(\phi)}{4}F_{\mu\nu}F^{\mu\nu}], \eqno{(2.1)}$$
where $\kappa^2=\frac{1}{M_{pl}^2}=8\pi G$, $\xi$ is the nonminimal coupling constant and $f(\phi)$ is the coupling function of the vector field. $F_{\mu\nu}$ is the gauge field strength given by $F_{\mu\nu}=\partial_\mu A_\nu - \partial_\nu A_\mu $. If $f=0$, the action reduces to the one of non-minimal Higgs inflation model, while if $\xi=0$, it reduces to the one studied in  \cite{soda1}.

We take $x$-axis to be the direction of the vector field without loss of generality and choose the gauge $A_0 =0$, thus $A_\mu =( 0, v(t) , 0, 0)$. Then we make the metric ansatz:
$$ds^2=-dt^2+e^{2\alpha(t)}(e^{-4\sigma(t)}dx^2+e^{2\sigma(t)}(dy^2+dz^2)), \eqno{(2.2)}$$
where $\dot{\alpha}$ represents the isotropic Hubble expansion rate and $\dot{\sigma}$ measures the anisotropic expansion rate.

The equations of motion can be obtained by doing variation of the action. Among them, $v(t)$ in the gauge field can be easily solved as
$$\dot{v}=f^{-2}(\phi)e^{-\alpha-4\sigma}p_A, \eqno{(2.3)}$$
with $p_A$ being an integration constant. Inserting Eq. $(2.3)$ into other equations, we obtain the background field equations
$$
\dot{\alpha}^2=\dot{\sigma}^2+\frac{\kappa^2}{3} \frac{1}{1+\kappa^2\xi\phi^2}[\frac{\dot{\phi}^2}{2}+\frac{\lambda}{4}\phi^4+\frac{f^{-2}(\phi)}{2}p_A^2e^{-4\alpha-4\sigma} -6\xi\phi\dot{\alpha}\dot{\phi}], \eqno{(2.4)}$$
$$
\ddot{\alpha}=-3\dot{\alpha}^2+\kappa^2 \frac{1}{1+\kappa^2\xi\phi^2}[\frac{\lambda}{4}\phi^4+\frac{1}{6}f^{-2}(\phi)p_A^2e^{-4\alpha-4\sigma} -(5\xi\phi\dot{\alpha}\dot{\phi}+\xi\dot{\phi}^2+\xi\phi\ddot{\phi})], \eqno{(2.5)}$$
$$
\ddot{\sigma}=-3\dot{\alpha}\dot{\sigma}+\frac{\kappa^2}{3} \frac{1}{1+\kappa^2\xi\phi^2} f^{-2}(\phi)p_A^2e^{-4\alpha-4\sigma} +\kappa^2 \frac{1}{1+\kappa^2\xi\phi^2} 2\xi\phi\dot{\sigma}\dot{\phi}, \eqno{(2.6)}$$
$$
\ddot{\phi}=-3\dot{\alpha}\dot{\phi}-\lambda \phi^3 +f^{-3}(\phi)\frac{df}{d\phi}p_A^2e^{-4\alpha-4\sigma} +6\xi\phi(\ddot{\alpha}+2\dot{\alpha}^2+\dot{\sigma}^2). \eqno{(2.7)}
$$

From Eq. $(2.4)$, the inflation in our model is determined by the effective potential $$V_{eff}=\frac{\lambda}{4}\phi^4+\frac{f^{-2}(\phi)}{2}p_A^2e^{-4\alpha-4\sigma},$$
where the second term comes from the vector contribution. It is convenient to define the ratio of the energy density of the vector field to that of the inflaton as
$$\Omega_A \equiv \frac{\rho_A}{\rho_\phi}=\frac{p_A^2f^{-2}(\phi)e^{-4\alpha-4\sigma}}{\lambda \phi^4/2}.$$
To be consistent with observational large-scale isotropy, $\Omega_A$ has to be small, i.e.  $\Omega_A \ll 1$. This requires that the potential of the inflaton is dominant during inflation. In other words, the early expansion of the universe is still driven by the inflaton potential. Consequently it is reasonable to work under slow-roll approximation: $\ddot{\phi}, \dot{\phi} \to 0$, with the slow-roll parameter $\epsilon_H\equiv -\frac{\dot{H}}{H^2} \ll 1$ and $\eta_H \equiv \epsilon_H-\frac{\ddot{H}}{2H\dot{H}} \ll 1$. One may also assume the spatial flatness with $\ddot{\alpha}+2\dot{\alpha}^2+\dot{\sigma}^2=\frac{R}{6} \to 0$.

On the other hand, if we want an unignorable anisotropy, the contribution from the vector field should not be diluted away completely by inflation. This suggests us to choose the function $f(\phi)$ carefully. For example, if we choose $f(\phi)=(\kappa\phi)^{-2}$, we have $\Omega_A=2p_A^2e^{-4\alpha-4\sigma}/\lambda$. During the inflation, the scale factor $e^{\alpha}$ grows quickly and $\Omega_A$ decays so fast leading to ignorable anisotropic contribution. Therefore, in order to have an unignorable anisotropy, it is reasonable to assume that
$$f\propto e^{-2c \alpha}  $$
with $c\geq 1$ to rebel the dilution due to the inflation. At the first looking, as $c \geq 1$, this may lead to fast increase of energy density of the vector field such that the anisotropy could be too large. However, this would not happen due to the existence of the attractor solution. Actually, the existence of the attractor solution is insensitive to $c$.

In the weak curvature coupling limit $\kappa^2\xi\phi^2 \ll 1$, our model reduces to the one studied in \cite{soda1}. In this case, it has been shown that when $f(\phi) \sim e^{-2c\alpha}, c\ge 1$, the energy density of the vector field grows during the inflation and cannot be neglected, leading to an anisotropic hair in the inflation. Consequently the inflation enters the second phase, the so-called anisotropic inflation, in which $\Omega_A$ reaches a sub-dominant but constant value. It turns out the anisotropic inflation phase is an attractor solution. Actually, it is more precise to use the metric components to characterize the anisotropy.  However, in the anisotropic inflation phase, $\Omega_A$ is proportional to $\dot{\sigma}/\dot{\alpha}$. Then the constancy of $\Omega_A$ indicates that the anisotropy cannot be diluted away, namely it increases with the scale factor.

Let us review briefly the anisotropy attractor solution in the weak curvature coupling limit, before our investigation into the strong curvature coupling limit of the model. We first work in the small anisotropic limit, that is, $\Omega_A$ can be negligible, in the first slow-roll phase. Under this assumption, the background metric  reduces to
$$ds^2=-dt^2+a^2(t)\delta_{ij}dx^idx^j, \eqno{(2.8)}$$
where $a(t)=e^{\alpha(t)}$. And the background field equations are now
$$H^2=\frac{\kappa^2}{3(1+\kappa^2\xi\phi^2)}[\frac{1}{2}\dot{\phi}^2+ \frac{\lambda}{4}\phi^4-6\xi H \phi \dot{\phi}], \eqno{(2.9)}$$
$$\ddot{\phi}+3H\dot{\phi}+\frac{\kappa^2\xi\phi^2(1+6\xi)}{1+\kappa^2\xi\phi^2(1+6\xi)}\frac{\dot{\phi}^2}{\phi}=-\frac{\lambda\phi^3}{1+\kappa^2\xi\phi^2(1+6\xi)}, \eqno{(2.10)}$$
where dots denote the derivatives with respect to the time $t$. In the case of $\kappa^2\xi\phi^2 \ll 1$, Eq. $(2.9)$ and Eq. $(2.10)$ reduce to
$$H^2=\frac{\kappa^2}{3}\frac{\lambda}{4}\phi^4, \eqno{(2.11)}$$
$$3H\dot{\phi}=-\lambda \phi^3. \eqno{(2.12)}$$
The slow-roll approximation requires that
$$\epsilon_H= \frac{8}{\kappa^2\phi^2} \ll 1, \hspace{3ex}\eta_H=  \frac{12}{\kappa^2\phi^2} \ll 1. \eqno{(2.13)}$$
Thus, the inflation occurs when $\kappa^2\phi^2 \gg 1$. As the totel number of e-foldings  $N \approx \frac{\kappa^2\phi^2}{8} \sim 60$, the coupling constant $\xi \ll 10^{-3}$.
Using Eq. $(2.11)$ and Eq. $(2.12)$, we obtain $\frac{d\alpha}{d\phi}=\frac{\dot{\alpha}}{\dot{\phi}}=-\frac{\kappa^2\phi}{4}$ and thus
$$f(\phi)=e^{\frac{c\kappa^2\phi^2}{4}}. \eqno{(2.14)}$$ 

When $c\ge 1$, the system can reach an attractive solution where the anisotropy reaches a sub-dominant ($\dot{\sigma}^2 \ll \dot{\alpha}^2$) but constant value corresponding to the second slow-roll stage. During this stage, the effect of the vector field in Eq. $(2.4)$ is still negligible even when it is comparable with that of the scalar field in Eq. $(2.7)$. The detailed analysis can be seen in \cite{soda1}. Thus the modified slow-roll equations are
$$H^2= \dot{\alpha}^2=\frac{\kappa^2\lambda\phi^4}{12}, \eqno{(2.15)}$$
$$3H\dot{\phi}= 3\dot{\alpha}\dot{\phi}=-\lambda \phi^3+\frac{c\kappa^2 p_A^2\phi}{2}e^{-4\alpha-4\sigma-c\kappa^2\phi^2/2}, \eqno{(2.16)}$$
where we have used Eq. $(2.14)$ and so
$$\frac{d\phi}{d\alpha}=-\frac{4}{\kappa^2\phi}+\frac{2cp_A^2}{\lambda\phi^3}e^{-4\alpha-4\sigma-c\kappa^2\phi^2/2}. \eqno{(2.17)}$$
Neglecting the evolution of $V(\phi)=\lambda\phi^4/4$, $dV/d\phi$ and $\sigma$, we obtain [see Appendix A for detail]
$$e^{4\alpha+4\sigma+c\kappa^2\phi^2/2}=\frac{c^2p_A^2}{c-1}\frac{\kappa^2}{2\lambda\phi^2}[1+\Omega e^{-4(c-1)\alpha+4\sigma}], \eqno{(2.18)}$$
with $\Omega=\frac{c-1}{c^2p_A^2}\frac{2\lambda\phi^2}{\kappa^2}\Omega_0e^{-4\sigma}$, $\Omega_0$ being the integration constant. Consequently, we have:
$$\frac{d\phi}{d\alpha}=-\frac{4}{\kappa^2\phi}+\frac{c-1}{c}\frac{4}{\kappa^2\phi}[1+\Omega e^{-4(c-1)\alpha+4\sigma}]^{-1}, \eqno{(2.19)}$$
and the following picture:

A) Initially $\alpha \to -\infty$, we have
$$\frac{d\phi}{d\alpha}=-\frac{4}{\kappa^2\phi}, \eqno{(2.20)}$$
corresponding to conventional isotropic slow-roll inflationary phase.

B) The second inflationary phase occurs when $\alpha \to \infty$
$$\frac{d\phi}{d\alpha}=-\frac{1}{c}\frac{4}{\kappa^2\phi}, \eqno{(2.21)}$$
$$e^{4\alpha+4\sigma+c\kappa^2\phi^2/2}=\frac{2c^2p_A^2}{c-1}\frac{\kappa^2}{4\lambda\phi^2}. \eqno{(2.22)}$$
Then $\Omega_A=\frac{1}{2}\frac{c-1}{c}\epsilon_H \equiv \frac{1}{2}I\epsilon_H$.

In this phase, Eq. $(2.4)$ and $(2.6)$ become
$$\dot{\alpha}^2=\frac{\kappa^2\lambda\phi^4}{12}, $$
$$3\dot{\alpha}\dot{\sigma}=\frac{\kappa^2}{3}f^{-2}(\phi)p_A^2e^{-4\alpha-4\sigma}, $$
where we have assumed $\ddot{\sigma} \ll \dot{\alpha}\dot{\sigma}$. And then
$$\frac{\dot{\sigma}}{\dot{\alpha}}=\frac{4}{3}\frac{f^{-2}(\phi)p_A^2e^{-4\alpha-4\sigma}}{\lambda\phi^4}=\frac{2}{3}\Omega_A. $$
Therefore we have $\dot{\sigma}/\dot{\alpha}= \frac{1}{3}I\epsilon_H$, corresponding to the anisotropic attractive solution we expect when $I \ge 0$.

\subsection{Strong curvature coupling limit}

Next we turn to the study of the anisotropic attractive solution in the strong curvature coupling limit. In the limit of $\kappa^2\xi\phi^2 \gg 1$, and in the slow-roll approximation $\dot{\phi} \to 0$ and $\dot{\phi}^2, \ddot{\phi} \ll H \dot{\phi}$,  Eq. $(2.9)$ and Eq. $(2.10)$ reduce to
$$H^2=\frac{\lambda \phi^2}{12 \xi}-2H\frac{\dot{\phi}}{\phi}, \eqno{(2.23)}$$
$$3H\dot{\phi}=-\frac{\lambda \phi}{\kappa^2 \xi (1+6\xi)}, \eqno{(2.24)}$$
Inserting Eq. $(2.24)$ back into Eq. $(2.23)$, we find that the second term on the right hand side of Eq. $(2.23)$ becomes $\frac{2\lambda}{3\kappa^2\xi(1+6\xi)}$, which could be neglected safely compared with the first term in the limit of $\kappa^2\xi\phi^2 \gg 1$. Thus Eq. $(2.23)$ becomes approximately
$$H^2=\frac{\lambda \phi^2}{12 \xi}. \eqno{(2.25)}$$
Then we find that  $\frac{d\alpha}{d\phi}=-\frac{\phi}{4}\kappa^2(1+6\xi)$ and $$f(\phi)=e^{\frac{c\kappa^2\phi^2}{4}(1+6\xi)},$$ with $c \ge 1$. In addition, it is straightforward to check that $\epsilon_H \ll 1$ and $\eta_H \ll 1$ in the strong coupling limit.

During anisotropic inflation, with similar analysis as in the weak coupling limit, the modified slow-roll equations become
$$H^2=\frac{\lambda \phi^2}{12 \xi}, \eqno{(2.26)}$$
$$3H\dot{\phi}=-\frac{\lambda \phi}{\kappa^2 \xi (1+6\xi)}+ \frac{c\kappa^2 p_A^2\phi}{2}e^{-4\alpha-4\sigma-(1+6\xi)c\kappa^2\phi^2/2},  \eqno{(2.27)}$$
 and then
$$\frac{d\phi}{d\alpha}=-\frac{4}{(1+6\xi)\kappa^2 \phi}+\frac{2cp_A^2\xi\kappa^2}{\lambda\phi}e^{-4\alpha-4\sigma-(1+6\xi)c\kappa^2\phi^2/2}. \eqno{(2.28)}$$
After integrating, we obtain [see Appendix A for detail]
$$\frac{d\phi}{d\alpha}=\frac{1}{1+6\xi}\{-\frac{4}{\kappa^2\phi}+\frac{c-1}{c}\frac{4}{\kappa^2\phi}[1+\Omega e^{-4(c-1)\alpha+4\sigma}]^{-1}\}, \eqno{(2.29)}$$
and have the anisotropic attractor solution when $\alpha \to \infty$
$$\frac{d\phi}{d\alpha}=-\frac{1}{c(1+6\xi)}\frac{4}{\kappa^2\phi}, \eqno{(2.30)}$$
$$e^{4\alpha+4\sigma+(1+6\xi)c\kappa^2\phi^2/2}=\frac{(1+6\xi)\xi c^2\kappa^4p_A^2}{2(c-1)\lambda}. \eqno{(2.31)}$$
In this case, Eq. $(2.4)$ and $(2.6)$ become
$$H^2=\frac{\lambda \phi^2}{12 \xi}, $$
$$3\dot{\alpha}\dot{\sigma}=\frac{f^{-2}(\phi)p_A^2e^{-4\alpha-4\sigma}}{3\xi\phi^2}+2\dot{\sigma}\frac{\dot{\phi}}{\phi}. $$
Using Eq. $(2.30)$, we have $\lvert \dot{\phi}/\dot{\alpha} \rvert \ll \phi$ in the strong curvature coupling limit. So then, $\lvert \dot{\phi}/\phi \rvert \ll \dot{\alpha}$ and the second term on the right hand side of the last relation can be neglected. Thus as in the weak curvature limit,
$$\frac{\dot{\sigma}}{\dot{\alpha}}=\frac{4}{3}\frac{f^{-2}(\phi)p_A^2e^{-4\alpha-4\sigma}}{\lambda\phi^4}=\frac{2}{3}\Omega_A.$$
However, now we find that
$$\Omega_A=\frac{1}{\kappa^2\xi\phi^2}I\epsilon_H, \eqno{(2.32)}$$
which is greatly suppressed in the strong curvature  coupling limit. Actually, from Eq. $(2.27)$, the effect of the vector field is suppressed by $e^{-\kappa^2\xi\phi^2}$ in the strong curvature coupling limit. Thus the suppression of $\Omega_A$ is reasonable.

\vspace{2em}More interestingly  this suppression behavior holds in more general cases when $f(\phi)=A(\kappa\phi)e^{\frac{c\kappa^2\phi^2}{4}(1+6\xi)}$, as long as the function $A(\kappa\phi)$ does not vary much during inflation. Similarly Eq. $(2.26)$ does not change while Eq. $(2.27)$ becomes
$$3H\dot{\phi}=-\frac{\lambda \phi}{\kappa^2 \xi (1+6\xi)}+ \frac{c\kappa^2 p_A^2\phi}{2A^2}e^{-4\alpha-4\sigma-(1+6\xi)c\kappa^2\phi^2/2},  \eqno{(2.33)}$$
where we have used the fact $\frac{A_{,\phi}}{A} \sim \frac{1}{\phi} \ll \frac{c}{2}(1+6\xi)\kappa^2\phi$. Thus,
$$\frac{d\phi}{d\alpha}=-\frac{4}{(1+6\xi)\kappa^2 \phi}+\frac{2cp_A^2\xi\kappa^2}{\lambda\phi A^2}e^{-4\alpha-4\sigma-(1+6\xi)c\kappa^2\phi^2/2}, \eqno{(2.34)}$$
which leads to
$$e^{4\alpha+4\sigma+(1+6\xi)c\kappa^2\phi^2/2}=\frac{(1+6\xi)\xi c^2\kappa^4p_A^2}{2(c-1)\lambda A^2}[1+\Omega e^{4(c-1)\alpha-4\sigma}]. \eqno{(2.35)} $$
Here we neglect  the evolution of $A(\kappa\phi)$ in the integration. When $\alpha \to \infty$, we have
$$\frac{d\phi}{d\alpha}=-\frac{1}{c(1+6\xi)}\frac{4}{\kappa^2\phi}, \eqno{(2.36)} $$
$$e^{4\alpha+4\sigma+(1+6\xi)c\kappa^2\phi^2/2}=\frac{(1+6\xi)\xi c^2\kappa^4p_A^2}{2(c-1)\lambda A^2}, \eqno{(2.37)} $$
corresponding to the anisotropic inflationary phase. And after straightforward calculation, we obtain
$$\Omega_A=\frac{1}{\kappa^2\xi\phi^2}I\epsilon_H, \eqno{(2.38)}  $$
exactly the same as Eq. $(2.32)$. Thus we come to the conclusion that the suppression of anisotropy is general in the strong curvature coupling limit. 

\subsection{Numerical analysis}

In the above discussion, we worked under approximations. In particular, we assume that the initial stage of the inflation is isotropic. It would be more convincing to support above pictures by numerical analysis. We can
 solve the Eqs. $(2.4)-(2.7)$ under different circumstances numerically. Here we pay more attention to the strong curvature coupling limit. We have set $\kappa^2 =1$ and chosen  $\xi=10^{9/2}$ and $\lambda =10^{-1}$. The reason for the choice of the parameters comes from our attempt to identify the inflaton with the Higgs boson in Standard Model of particle physics. On the one hand, the data on the CMB anisotropies gives the constraint $\lambda/\xi^2 \simeq 10^{-10}$\citet{higgs1,komatsu1}. On the other hand, from the properties of the Higgs boson, $\lambda \sim M_H^2/\nu^2$, with the mass of Higgs boson $M_H \approx 125$ GeV\citet{lhc1,lhc2}  and the vacuum expectation value of the Higgs field $\nu \approx 246$ GeV. Thus, to identify the Higgs boson with the inflaton, $\lambda$ should be of order $10^{-1}$ and then $\xi \sim 10^{9/2}$. Additionally, as the e-folding number $N \approx \frac{1+6\xi}{8\xi}\kappa^2\xi\phi^2(0)$\citet{komatsu1} in the isotropic case, we choose the initial value $\phi(0) = 0.05$.

 In Figure 1, we depict the evolution of $\phi - \dot{\phi}$ in the strong curvature coupling limit, where $f(\phi)=e^{(6\xi+1) \kappa^2 \phi^2 /2}$. We have chosen $\phi(0) = 0.05$ and $\dot{\phi}(0)=0$. From the figure, we can see clearly that there are two slow-roll inflationary phases. Moreover, we find that $\phi = 0.01 \sim 0.03$ during the anisotropic inflation so that the suppression factor is about $10^{-1.5}$ from Eq. (2.32), which agrees with numerical result (red thick line in Fig. 2) nicely.

\begin{figure}[H]
\centering
\renewcommand{\captionfont}{\small}
\setcaptionwidth{0.8\textwidth}
\includegraphics[width=0.7\textwidth,height=0.4\textheight]{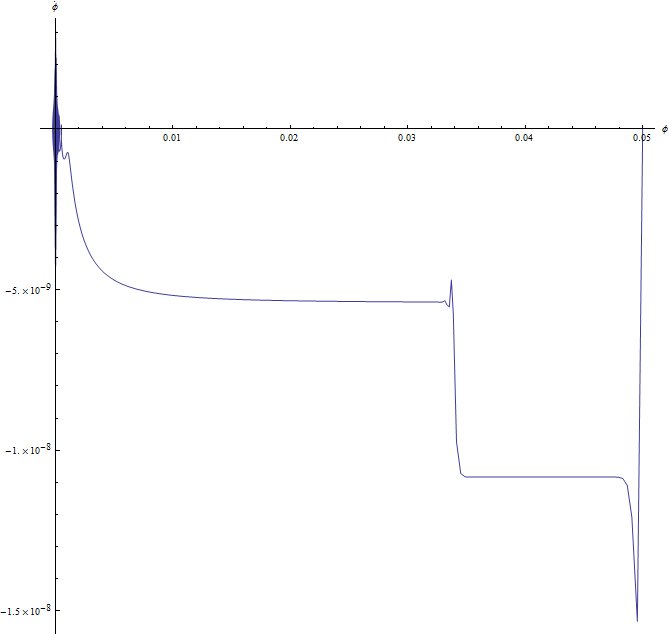}
\caption{Phase flow for $\phi$ in the strong curvature coupling limit with $\xi=10^{9/2}$, $\lambda =10^{-1}$ and the initial conditions $\phi(0)=0.05$ and $\dot{\phi}(0)=0$. Two different slow-roll phases can be observed clearly.}
\end{figure}

In Figure 2,  the evolutions of the anisotropy $\dot{\sigma}/\dot{\alpha}$  in different cases are displayed. The initial condition is set as $\dot{\sigma}/\dot{\alpha}|_{t=0}=0$. The horizontal axis is chosen to be  the e-folding number $N$. When the coupling function $f(\phi)$ includes an exponentially dependent form, the blue dashed line corresponding to the weak curvature coupling limit while the red thick line and the black line corresponding to the strong curvature coupling limit, the anisotropy grows rapidly in the first slow-roll stage and then behaves like an attractor in the second stage, which is exactly the anisotropic inflation we expect. Moreover in the strong curvature coupling limit, the anisotropy is suppressed greatly. This fact can be seen clearly after zooming in the details in the anisotropic inflation phase. Compared with the one in the weak curvature coupling limit, the anisotropy in the strong curvature coupling limit is suppressed by a factor of order $10^{1.5}$, as can be read from the zoomed-in diagram. Furthermore, Fig. 2 shows the existence of attractor solution and the suppressed anisotropy even for other choices of the coupling function $f(\phi)$  as long as the function  $f(\phi)$ is proportional to $e^{\frac{c\kappa^2\phi^2}{4}(1+6\xi)}$.
 In contrast, when the coupling function is of a power-law form, corresponding to the green line in the figure, the anisotropy will be damped quickly.

\begin{figure}
\captionstyle{flushleft}
\hspace{2em}
\begin{minipage}[t]{0.5\linewidth}
\centering
\renewcommand{\captionfont}{\scriptsize}
\setcaptionwidth{\linewidth}
\includegraphics[width=3 in, height=1.9 in]{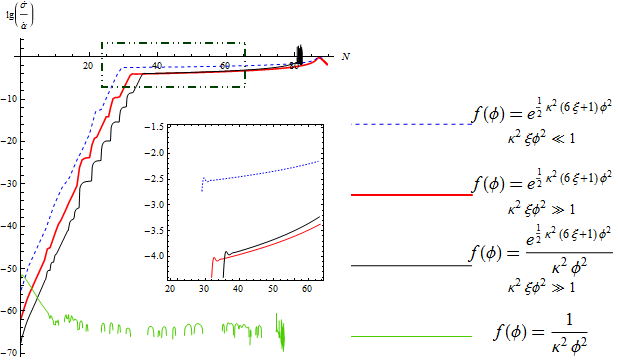}
\caption{Evolution of the anisotropy $\dot{\sigma}/\dot{\alpha}$ under different choices of the coupling function and the parameters, with the initial condition $\dot{\sigma}/\dot{\alpha}|_{t=0}=0$. The anisotropy is suppressed by about $10^{1.5}$ in the strong curvature coupling limit $\kappa^2\xi\phi^2 \gg 1$, where we have set $\xi =10^{9/2}$ and $\lambda =10^{-1}$. The box diagram in the middle shows finer details of different evolutions in the anisotropic inflation, after zooming in.}
\label{Fig. 2}
\end{minipage}%
\hspace{0.5em}
\begin{minipage}[t]{0.37\linewidth}
\centering
\renewcommand{\captionfont}{\scriptsize}
\setcaptionwidth{\linewidth}
\includegraphics[width=2 in, height=1.9 in]{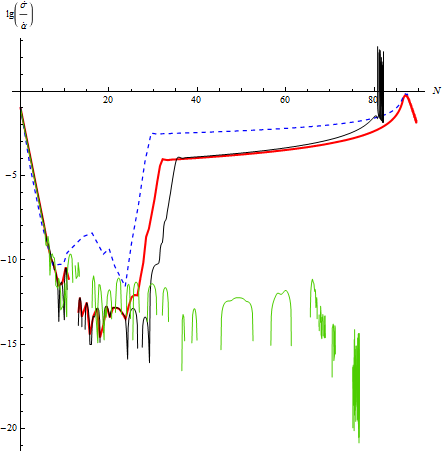}
\caption{Evolution of the anisotropy $\dot{\sigma}/\dot{\alpha}$, with the initial condition $\dot{\sigma}/\dot{\alpha}|_{t=0}=10^{-1}$. The coupling function and the parameters are chosen to be the same as in Fig. 2. The initial anisotropy damped quickly in the first  few e-folding numbers. }
\label{Fig. 3}
\end{minipage}
\end{figure}


In Figure 3, we calculate the evolution of the anisotropy $\dot{\sigma}/\dot{\alpha}$ even when it is relatively large initially. The functions and the parameters are chosen to be the same as in Fig. 2. In contrast, here we set $\dot{\sigma}/\dot{\alpha}|_{t=0}=10^{-1}$. Such a large initial anisotropy damped quickly in the first slow-roll stage and then evolved as in Fig. 2. Thus the assumption of isotropic inflation in the first slow-roll phase in our analytic analysis is justified.




\section{Conclusion}

In this work, we studied a new anisotropic inflation model, with the inflaton being nonminimally coupled with the gravity and a vector field simultaneously. We found the anisotropic attractor solutions in both the weak and strong curvature coupling limit and  that the contribution from the anisotropy is greatly suppressed in the strong curvature coupling limit.

One remarkable feature of this kind of model is that at the strong curvature  coupling limit, the anisotropic energy density is suppressed by a factor $1/\kappa^2\xi\phi^2$. And such a suppression happens for a generic form of coupling function. 
However, the recent study in \cite{Obata:2014qba} suggests there could be nontrivial dynamics for electroweak gauge field
during the Higgs inflation. It would be interesting to investigate this issue more carefully.

\vspace{1em}\noindent {\large{\bf Acknowledgments}}\\

 The work was in part supported by NSFC Grant No.~11275010, No.~11335012 and No.~11325522.
\vspace*{5mm}

\section*{Appendix A}

Here we give the solution to the differential equation
$$\frac{d\phi}{d\alpha}=-\frac{V_\phi}{\kappa^2 V}+2c\frac{p_A^2}{V_\phi}e^{-4\alpha-4\sigma-4c\kappa^2\int\frac{V}{V_\phi}d\phi}, \eqno {(A.1)}$$
 where $V_\phi \equiv dV/d\phi$. Eq. $(2.17)$ is just a particular example of Eq. $(A.1)$ when $V(\phi)=\lambda\phi^4/4$.

Rearranging Eq. $(A.1)$, we have
$$4c\kappa^2\frac{V}{V_\phi}e^{4c\kappa^2\int\frac{V}{V_\phi}d\phi}(\frac{d\phi}{d\alpha}+\frac{V_\phi}{\kappa^2 V})=\frac{8c^2\kappa^2 p_A^2 V}{V_\phi^2}e^{-4\alpha-4\sigma}. \eqno{(A.2)}$$
Defining $e^{4c\kappa^2\int\frac{V}{V_\phi}d\phi} \equiv F(\alpha)$ and setting $F(\alpha)=g(\alpha)e^{-4\alpha-4\sigma}$, we find that  Eq. $(A.2)$ reduces to
$$\frac{dg(\alpha)}{d\alpha}+4(c-1)g(\alpha) = \frac{8c^2\kappa^2 p_A^2 V}{V_\phi^2}, \eqno{(A.3)}$$
where we have neglected the evolution of $\sigma$. As the inflation occurs in the slow-roll regime, the evolution of $V$ and $V_\phi$ can be neglected.  We finally get
$$g(\alpha)=e^{4\alpha+4\sigma+4c\kappa^2\int\frac{V}{V_\phi}d\phi}=\frac{2c^2\kappa^2p_A^2 V}{(c-1)V_\phi^2}[1+\Omega e^{4(1-c)\alpha+4\sigma}], \eqno{(A.4)}$$
where $\Omega=\frac{(c-1)V_\phi^2\Omega_0 e^{-4\sigma}}{2c^2\kappa^2p_A^2 V}$, with $\Omega_0$ being the integration constant in solving Eq. $(A.3)$.

Setting $V(\phi)=\lambda\phi^4/4$, we have Eq. $(2.18)$. We obtain Eq. $(2.19)$ by inserting Eq. $(2.18)$ back into Eq. $(2.17)$.

\vspace{2em}Eq. $(2.29)$ and Eq. $(2.35)$ can be obtained in a similar way, but the details are slightly different. To obtain Eq. $(2.29)$, we first rearrange Eq. $(2.28)$ as
$$\frac{d\phi}{d\alpha} (1+6\xi) c \kappa^2\phi +4c=\frac{2c^2p_A^2\xi\kappa^4(1+6\xi)}{\lambda} e^{-4\alpha-4\sigma-(1+6\xi)c\kappa^2\phi^2/2}. \eqno{(A.5)}$$
Defining $f(\alpha) \equiv  (1+6\xi)c\kappa^2\phi^2/2$ and multiplying $e^{f(\alpha)}$ on both sides of Eq. $(A.5)$, we have
$$\frac{e^{f(\alpha)}}{d\alpha}+4ce^{f(\alpha)}=\frac{2c^2p_A^2\xi\kappa^4(1+6\xi)}{\lambda}e^{-4\alpha-4\sigma}. \eqno{(A.6)}$$
Setting $e^{f(\alpha)} = g(\alpha)e^{-4\alpha-4\sigma}$, we read Eq. $(A.6)$
$$\frac{dg(\alpha)}{d\alpha}+4(c-1)g(\alpha)=\frac{2c^2p_A^2\xi\kappa^4(1+6\xi)}{\lambda}, \eqno{(A.7)}$$
where we have neglected the evolution of $\sigma$.
Solving this inhomogeneous ordinary differential equation, we obtain
$$g(\alpha)=e^{4\alpha+4\sigma+(1+6\xi)c\kappa^2\phi^2/2}=\frac{(1+6\xi)c^2p_A^2\xi\kappa^4}{2(c-1)\lambda}[1+\Omega e^{4(1-c)\alpha+4\sigma}], \eqno{(A.8)}$$
where $\Omega=\frac{2(c-1)\lambda\Omega_0e^{-4\sigma}}{(1+6\xi)c^2p_A^2\xi\kappa^4}$, with $\Omega_0$ being the integration constant in solving Eq. $(A.7)$.
Inserting Eq. $(A.8)$ back into Eq. $(2.28)$, we finally get Eq.$(2.29)$.


\end{CJK*}
\end{document}